\documentclass[prb,aps,twocolumn]{revtex4-1}
\usepackage{bm}
\usepackage{amsmath}
\usepackage{amssymb} 
\usepackage{epsfig}
\usepackage{graphicx}
\usepackage{color}
\usepackage{xcolor}
\usepackage{ulem}
\usepackage{cancel}
\usepackage{hyperref}
\newcommand{\pderiv}[2]{\frac{\partial #1}{\partial #2}}
\renewcommand{\phi}{\varphi}
\renewcommand{\kappa}{\varkappa}

\renewcommand{\i}{\mathrm i}
\newcommand{\e}{\mathrm e}
\newcommand{\eps}{\varepsilon}

\newcommand{\aver}[1]{\left \langle #1 \right \rangle}

\begin{document}

\title{Photovoltaic Hall effect in two-dimensional electron gas: Kinetic theory}

\author{M.\,V.\,Durnev}

\affiliation{Ioffe Institute, 194021 St.\,Petersburg, Russia}

\begin{abstract}
We study theoretically transverse photoconductivity induced by circularly polarized radiation, i.e. the photovoltaic Hall effect, and linearly polarized radiation causing intraband optical transitions in two-dimensional electron gas (2DEG). 
We develop a microscopic theory of these effects based on analytical solution of the Boltzmann equation for arbitrary electron spectrum and scattering mechanism. We calculate the transverse photoconductivity of 2DEG with parabolic and linear dispersion for short-range and Coulomb scatterers at different temperatures. We show that the transverse electric current is significantly enhanced at frequencies comparable to the inverse energy relaxation time, whereas at higher frequencies the excitation spectrum and the direction of current depend on the scattering mechanism. We also analyse the effect of thermalization processes caused by electron-electron collisions on the photoconductivity.
\end{abstract}

\maketitle

\section{Introduction}

Direct electric currents induced by high-frequency electric field in graphene and other two-dimensional (2D) materials are subject of active investigation in recent years~\cite{Glazov2014, Koppens2014}. In unbiased samples, the dc current is induced by oscillating electric and magnetic fields of the incident radiation through different mechanisms, including photothermoelectric~\cite{Xu2010, Cai2014, Castilla2019}, photovoltaic and bolometric~\cite{Freitag2013}, plasmonic~\cite{Vicarelli2012,Muraviev2013, Bandurin2018}, photon drag~\cite{Karch2010,Entin2010,Obraztsov2014},  bulk~\cite{Tarasenko2011,Drexler2013, Geller2015, Fateev2017, Kheirabadi2018,Quereda2018, Ma2019} and edge~\cite{Karch2011, Candussio2020, Durnev2021} photogalvanic effects. These mechanisms are relevant in 2D structures with broken space inversion symmetry --  due to crystal lattice, $p$-$n$ junctions, inhomogeneity of illumination, photon wave vector or edges. On the other hand, in a biased 2D layer, incident radiation may induce dc electric current, which flows in the direction perpendicular to the dc electric field, even if the system is spatialy isotropic. For circularly polarized radiation, the appearance of such a transverse current is reminiscent of the Hall effect, and hence, is termed as the photovoltaic Hall effect~\cite{Oka2009}.

The photo-induced anisotropy of conductivity was studied in early works in three-dimensional (3D) crystals and thin films~\cite{Galpern1969, Belinicher1981, Karaman1983, Esayan1984}. It was shown that under illumination of a biased 3D crystal with circularly or linearly polarized electromagnetic wave the transverse dc current appears, which direction is controlled by the radiation polarization. The interest in transverse photoconductivity has been recently renewed with an advent of graphene and 2D materials. The photovoltaic Hall effect and transverse photoconductivity induced by linearly polarized radiation in 2D layers are being actively studied both theoretically~\cite{Oka2009, Zavyalov2010, Trushin2011, Sato2019} and experimentally~\cite{Yin2011, Seifert2019, McIver2020}. 

The anisotropic photoconductivity in 2D materials has been studied so far for high-frequency radiation, which induces interband optical transitions between the valence and conduction bands in low-intensity regime~\cite{Trushin2011, Sato2019}, and opens Floquet gaps in high-intensity regime~\cite{Oka2009, Morina2015, Kristinsson2016}. With decreasing the frequency of radiation, e.g. to the terahertz range, the interband transitions in doped structures become forbidden and intraband (Drude-like) optical transitions in two-dimensional electron gas (2DEG) come to play. The anisotropic photoconductivity of 2DEG in the low-frequency regime has been studied numerically~\cite{Zavyalov2010}, whereas the analytical theory is missing. 

Here, we study the photovoltaic Hall effect and transverse photoconductivity induced by linearly polarized radiation caused by intraband optical transitions in 2DEG.  The transverse electric current contains two contributions: The first one is due to alignment of electron momenta by ac and static electric fields, and the second one is due to the dynamic heating of 2DEG.  We develop a microscopic theory of anisotropic photoconductivity based on solution of the Boltzmann equation for an arbitrary momentum dependence of the electron energy and arbitrary scattering mechanism. It allows one to apply the developed theory to a large class of 2D materials with linear, parabolic or Dirac energy dispersion, such as monolayer and bi-layer graphene, monolayers of TMDC and quantum wells.  We show that the ``heating'' contribution is dominant in the frequency range $\omega \tau_0 \lesssim 1$, where $\omega$ is the radiation frequency and $\tau_0$ is the energy relaxation time. In this range, the transverse photoconductivity reaches $\sim 1$~\% of  the ``dark'' conductivity of 2DEG at 1 W cm$^{-2}$ of the radiation intensity. At  $\omega \tau_0 \gg 1$ the transverse photoconductivity is determined by the relaxation times of the first and second angular harmonics of the distribution function and their energy derivatives. In this case, the excitation spectrum and even the sign of the transverse current are governed by the prevalent scattering mechanism. We also show that thermalization process caused by electron-electron collisions has a low impact on the photoconductivity at low temperatures, however
may considerably alter the photoconductivity excitation spectrum with increasing temperature. 

The paper is organized as follows. In Sec.~\ref{model} we formulate the model and present kinetic equations. In Sec.~\ref{parabolic} we calculate transverse photoconductivity of 2DEG with parabolic energy dispersion. In Sec.~\ref{arbitrary} we calculate transverse photoconductivity of 2DEG with an arbitrary energy dispersion and apply the derived results to analyze the photoconductivity of graphene. In Sec.~\ref{thermal} we study the role of thermalization process on the photoconductivity, and in Sec.~\ref{conclusion} we summarize the results.

\section{Model and kinetic equations} \label{model}

\begin{figure}[htpb]
\includegraphics[width=0.49\textwidth]{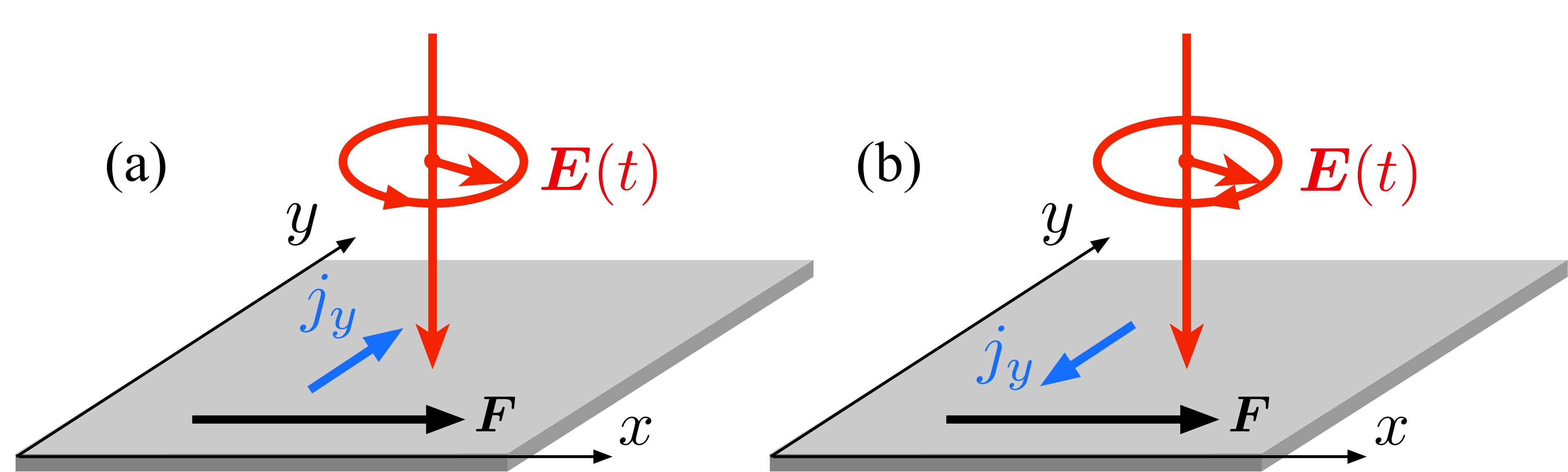}
\caption{\label{fig1} The sketch of the photovoltaic Hall effect. Circularly polarized radiation induces transverse dc current, which direction is opposite for the right-handed (a) and left-handed (b) circular polarization. 
}
\end{figure}

We consider 2DEG in the $(xy)$ plane subject to in-plane oscillating electric field $\bm E(t)$ and static electric field $\bm F$. In addition to the current along $\bm F$, the oscillating and static fields create the direct electric current, which flows in the direction perpendicular to $\bm F$. This transverse current appears for circularly and linearly polarized field $\bm E(t)$
and changes its direction when the polarization sign is changed, see Fig.~\ref{fig1}. 

In general, the photo-induced dc current in isotropic 2DEG is determined by three constants $\gamma_j$~\cite{Belinicher1981}
\begin{multline}
\label{symmetry}
\bm j = \gamma_{1} |\bm E|^2\bm F + \gamma_2 \left[ \bm E^* (\bm E \cdot \bm F) + \mathrm{c.c} \right]
\\ + \i \gamma_3 \left[ \bm F \times \left[ \bm E \times \bm E^* \right] \right]\:.
\end{multline}
Here, $\gamma_1$ describes the change of isotropic conductivity under the action of radiation, whereas $\gamma_2$ and $\gamma_3$ describe the anisotropic photoconductivity induced by linearly and circularly polarized radiation, respectively.  Below we calculate $\gamma_2$ and $\gamma_3$ due to intraband (Drude-like) optical transitions in 2DEG. Note, that Eq.~\eqref{symmetry} with no additional parameters also holds for 2DEG in high-symmetry 2D crystals, such as graphene.

To calculate the transverse current we introduce the electron distribution function in the momentum space $f(\bm p, t)$, which satisfies the Boltzmann equation
\begin{equation}
\label{kinetic}
\pderiv{f}{t} + e [\bm F + \bm E(t)] \cdot\pderiv{f}{\bm p} = \mathrm{St}~f\:.
\end{equation} 
Here $\bm p$ is the electron momentum, $e$ is the electron charge, $\bm E(t) = \bm E \exp(-\i\omega t) + \mathrm{c.c}$, the static electric field $\bm F$ is parallel to the $x$-axis, and $\mathrm{St}~f$ is the collision integral. Equation~\eqref{kinetic} is valid for the classical regime, when $\hbar \omega \ll \bar{\eps}$, where $\bar{\eps}$ is the mean electron energy.

We solve Eq.~\eqref{kinetic} analytically by expanding the distribution function $f(\bm p, t)$ in the series in the electric field amplitude as follows
\begin{multline}
\label{expand}
f(\bm p, t) = f_0 + \bar{f_{1}}(\bm p) + [ \tilde{f_{1}}(\bm p) \e^{-\i\omega t} + \mathrm{c.c.}] \\
+ \bar{f_{2}}(\bm p)  + [\tilde{f_{2}}(\bm p) \e^{-\i\omega t} + \mathrm{c.c.}] + \bar{f_3}(\bm p)\:.
\end{multline} 
In the absence of electric field, the electron distribution is equilibrium and described by the Fermi-Dirac function $f_0$ with chemical potential $\mu_0$ and temperature $T_0$. The first order corrections $\bar{f_1} \propto F$ and $\tilde{f_1} \propto E$ determine linear (Drude) conductivity, responsible for dc electric current and ac current oscillating at the field frequency, respectively. The second-order corrections $\bar{f_2} \propto EE^*$ and $\tilde{f_2} \propto FE$. The transverse dc current is determined by the third-order correction $\bar{f_3} \propto FEE^*$. Note that we do not consider second-order corrections $\propto F^2$ and $\propto E^2$, since these corrections do not contribute to the Hall current. Considering the term $e [\bm F + \bm E(t)] \cdot \partial f/ \partial \bm p$ in Eq.~\eqref{kinetic} as a perturbation, we obtain the following equations for corrections to distribution function:
\begin{subequations}
\label{kinetic2}
\begin{align}
\label{barf1}
e \bm F \cdot \pderiv{f_0}{\bm p} = \mathrm{St}~\bar{f_1}\:,  \\
\label{tildef1}
-\i\omega \tilde{f_1} + e \bm E \cdot \pderiv{f_0}{\bm p} = \mathrm{St}~\tilde{f_1}\:,  \\
\label{barf2}
e \left( \bm E^* \cdot \pderiv{\tilde{f_1}}{\bm p} +  \bm E \cdot \pderiv{\tilde{f_1^*}}{\bm p} \right) = \mathrm{St}~\bar{f_2}\:,  \\
\label{tildef2}
-\i\omega \tilde{f_2} + e \left( \bm F \cdot \pderiv{\tilde{f_1}}{\bm p} +  \bm E \cdot \pderiv{\bar{f_1}}{\bm p} \right) = \mathrm{St}~\tilde{f_2}\:,  \\
\label{barf3}
e \bm F \cdot \pderiv{\bar{f_2}}{\bm p}  +e \left( \bm E^* \cdot \pderiv{\tilde{f_2}}{\bm p} + \bm E \cdot \pderiv{\tilde{f_2^*}}{\bm p} \right) = \mathrm{St}~\bar{f_3}\:. 
\end{align}
\end{subequations}

We use the relaxation time approximation for the collision integral. The relaxation of the first and second angular harmonics of the function $f(\bm p, t)$ is described by the times $\tau_1$ and $\tau_2$ defined as $\tau_1^{-1} = - \aver{\bm v \mathrm{St}~f}/\aver{\bm v f}$ and  $\tau_2^{-1} = - \aver{v_x v_y \mathrm{St}~f}/\aver{v_x v_y f}$, respectively, where $\bm v = \partial \eps/\partial \bm p$ is the electron velocity, $\eps$ is energy, and the angular brackets denote averaging over $\bm p$ directions. As shown below, the zeroth angular harmonic of $f(\bm p,t)$ also contributes to the transverse current. To describe the relaxation of the zeroth angular harmonic $\aver{f(\bm p, t)}$ we use the following collision integral:
\begin{equation}
\label{St_zero}
\mathrm{St}~\aver{f} = -\frac{\aver{f} - f_{T}(\mu,T)}{\tau_{ee}} - \frac{\aver{f} - f_0}{\tau_0}\:.
\end{equation}
The first term in the right-hand side of Eq.~\eqref{St_zero} describes thermalization of the electron distribution towards the Dirac-Fermi distribution $f_{T}$ caused by electron-electron collisions. The $f_T$ distribution is characterized by chemical potential $\mu$ and temperature $T$ defined from the conservation of electron number and total energy $\sum_{\bm p} f = \sum_{\bm p} f_{T}$ and $\sum_{\bm p} \eps f = \sum_{\bm p} \eps f_{T}$, where $\eps$ is the electron energy. The thermalization is determined by the time $\tau_{ee}$. The second term in the right-hand side of Eq.~\eqref{St_zero} describes the energy relaxation by phonons determined by the time $\tau_0$. In what follows we consider $\tau_{ee}$ and $\tau_0$ independent of energy, but, generally, $\tau_0$ and $\tau_{ee}$ depend on temperature. Further, we consider a temperature range where $\tau_{1,2} \ll \tau_{ee} \ll \tau_0$. This inequality is relevant for 2DEG at low temperatures, however may also hold at higher temperatures, if the phonon-assisted energy relaxation is suppressed, as in the case of graphene~\cite{Strait2011, Song2012}.

The transverse current is determined by $\bar{f_3}$ and given by
\begin{equation}
\label{jy}
j_y = e g \sum \limits_{\bm p} v_y \bar{f_3}\:,
\end{equation} 
where $g$ is the factor of spin and valley degeneracy. 
Multiplying Eq.~\eqref{barf3} by $v_y$ and averaging the result over the directions of $\bm p$, we obtain
\begin{equation}
\label{f3aver}
\aver{v_y \bar{f_3}} = -e \tau_1 \aver{v_y  \left( \bm F \cdot \pderiv{\bar{f_2}}{\bm p}  + \bm E^* \cdot \pderiv{\tilde{f_2}}{\bm p} + \bm E \cdot \pderiv{\tilde{f_2^*}}{\bm p} \right)}\:.
\end{equation}
Summation of Eq.~\eqref{f3aver} over $\bm p$ and integration by parts yield
\begin{equation}
\label{jy_3}
j_y = e^2 g \sum \limits_{\bm p} \left(\bar{f_2} \bm F  + \tilde{f_2} \bm E^* +  \tilde{f_2^*} \bm E \right) \cdot \pderiv{(v_y \tau_1)}{\bm p} \:.
\end{equation} 
Equation~\eqref{jy_3} is used further to calculate transverse current for 2DEG with arbitrary dispersion $\eps(|\bm p|)$.

\section{Electron gas with parabolic dispersion} \label{parabolic}

We start with calculating $j_y$ for parabolic energy dispersion of electrons $\eps(p) = p^2/2m$, where $m$ is the effective mass, and $p = |\bm p|$. This important limit also helps to clarify the basic physics of the effect. Calculating derivative in the right-hand side of Eq.~\eqref{jy_3} one obtains
\begin{multline}
\label{jy_par1}
j_y = e^2 g F\sum \limits_{\bm p} v_x v_y \tau_1' \bar{f_2} + \frac{e^2 g}{m}  \sum \limits_{\bm p} (\tau_1 \eps)' \left( E_y^* \tilde{f_2} + E_y \tilde{f_2^*} \right)
\\
+ \frac{e^2 g}{2} \sum \limits_{\bm p}  \left[ 2 v_x v_y E_x^* \tilde{f_2} - (v_x^2 - v_y^2) E_y^*\tilde{f_2} + \mathrm{c.c}  \right] \tau_1'  \:.
\end{multline} 
Here $(...)' = \partial (...)/\partial \eps$, and we took into account that $\bm F \parallel x$. 
Electric current~\eqref{jy_par1} contains three contributions. The first one, proportional to $v_x v_y \bar{f_2}$, is related to the optical alignment of electron momenta by the oscillating electric field~\cite{Zemskii1976, Dymnikov1976, Merkulov1990, Hartmann2011, Golub2011}. Optical alignment results in excess electrons and holes below Fermi level with velocities directed at $\pi/4$ and $3\pi/4$ with respect to the $x$-axis. Together with the static electric field, which causes imbalance between charge carriers with $v_x > 0$ and $v_x < 0$, it results in a net $y$ current. The second term in Eq.~\eqref{jy_par1}, proportional to $(\tau_1 \eps)' \tilde{f_2}$, is related to the dynamic heating and cooling of 2DEG by the combined action of static and oscillating fields~\cite{Galpern1969}. Electrons periodically gain and lose their energy with a rate proportional to $(\bm F \cdot \bm E)$ and oscillating in time at frequency $\omega$. In turn, $E_y^*$ component of the incident field drives the electrons along the $y$-direction. At the first and second half-periods of an oscillation cycle, both distribution of velocities and momentum relaxation times of electrons are different, which results in a net electric current. Finally, the third contribution to electric current~\eqref{jy_par1} is related to dynamic optical alignment of charge carrier momenta by combined action of $\bm E(t)$ and $\bm F$. The first and second contributions in Eq.~\eqref{jy_par1} yield transverse photoconductivity for linearly polarized radiation with nonzero $E_x E_y^* + E_x^* E_y$, whereas the second and third contributions result in the photovoltaic Hall effect.

Solutions of Eqs.~\eqref{barf1}, \eqref{tildef1} for the first order corrections to distribution function are 
\begin{align}
\label{f1}
\bar{f_1} &= -e \tau_1 (\bm F \cdot \bm v) f_0'\:, \\
\tilde{f_1} &= -e \tau_{1\omega} (\bm E \cdot \bm v) f_0'\:, \nonumber
\end{align}
where $\tau_{1\omega} = \tau_{1}/(1 - \i \omega \tau_1)$. Substituting Eq.~\eqref{f1} in Eqs.~\eqref{barf2} and \eqref{tildef2} and solving the kinetic equations we obtain for the second-order corrections
\begin{eqnarray}
\label{f2}
\bar{f_2} &=& \aver{\bar{f_2}} + e^2 |\bm E|^2 \tau_2 \left(\mathrm{Re}\{\tau_{1\omega}\} f_0' \right)' \left[S_1 (v_x^2 - v_y^2)+ 2S_2 v_x v_y \right] \:, \nonumber \\
%\aver{\bar{f_2}} &=& f_{\rm eq}(\bar{\mu},\bar{T}) - f_0 + \frac{e^2|\bm E|^2 \tau_{ee}}{m} \left( \tau_{1\omega} f_0' \eps \right)'  + \mathrm{c.c.} \nonumber \\
\tilde{f_2} &=& \aver{\tilde{f_2}} + \frac{e^2 F \tau_{2\omega} \left[( \tau_{1\omega} + \tau_1) f_0' \right ]' }{2} \times \nonumber \\
&\times& \left[E_x (v_x^2 - v_y^2) + 2E_y v_x v_y \right]\:,
%\aver{\tilde{f_2}} &=& f_{\rm eq}(\tilde{\mu},\tilde{T}) - f_0 + \frac{e^2(\bm E \cdot \bm F) \tau_{ee}}{m} \left[(\tau_{1\omega} + \tau_1) f_0' \eps \right]'  + \mathrm{c.c.} 
\end{eqnarray}
where $S_1 = (|E_x|^2 - |E_y|^2)/|\bm E|^2$ and $S_2 = (E_xE_y^* + E_x^* E_y)/|\bm E|^2$ are the Stokes parameters of the radiation polarization, $\tau_{2\omega} = \tau_2/(1 - \i \omega \tau_2)$, and $\mathrm{Re}$ stands for real part. As seen from Eq.~\eqref{f2}, the functions $\bar{f_2}$ and $\tilde{f_2}$ are sums of the zeroth and second angular harmonics in momentum space. Further in this section, we neglect thermalization by setting $\tau_{ee} \to \infty$ in Eq.~\eqref{St_zero}, and hence find $\langle \tilde{f_2} \rangle = -e \tau_{0\omega} \langle \bm F \cdot \partial \tilde{f_1}/\partial \bm p +  \bm E \cdot \partial \bar{f_1}/ \partial \bm p \rangle$ from Eq.~\eqref{tildef2}. As shown in  Sec.~\ref{thermal}, the neglect of thermalization is eligible at low temperatures of 2DEG.

Substitution of Eq.~\eqref{f2} in Eq.~\eqref{jy_par1} for the current, averaging over $\bm p$ directions and integration by parts yield
\begin{multline}
\label{jy_par2}
j_y = - \frac{e^4 g F |\bm E|^2}{m^2} \times \\
\left \{S_2 \mathrm{Re} \sum \limits_{\bm p} \left[ \left( \eps^2 \tau_1' \tau_2\right)' \tau_{1\omega} + \eps (\tau_1 \eps)'' \tau_{0\omega}(\tau_{1\omega} + \tau_1) \right] f_0' \right.\\ \left.
-  S_3 \mathrm{Im} \sum \limits_{\bm p}  \left[ \left( \eps^2 \tau_1' \tau_{2\omega} \right)' - \eps (\tau_1 \eps)'' \tau_{0\omega} \right] (\tau_{1\omega} + \tau_1)  f_0' \right\}\:,
\end{multline} 
where $\tau_{0\omega} = \tau_0/(1 - \i \omega \tau_0)$, and $S_3 = \i(E_x E_y^* - E_x^* E_y)/|\bm E|^2$ is the degree of circular polarization. 
Finally, Eqs.~\eqref{symmetry} and \eqref{jy_par2} yield at $T_0 = 0$ 
\begin{align}
\label{gamma_par}
\gamma_2 &= \frac{\sigma_0 e^2}{m} \mathrm{Re} \left\{ (\eps_F \tau_1'' + 2\tau_1')\alpha_\omega\tau_{0\omega} \right. \nonumber  \\
& ~~~~~~~~~~~~~~~~\left. + (1-\i\omega\tau_1)^{-1} [2 \tau_1' \tau_2 + \eps_F(\tau_1' \tau_2)'] \right\}\:, \nonumber  \\
\gamma_3 &= \frac{\sigma_0 e^2}{m} \mathrm{Im}\left \{ 
(\eps_F \tau_1'' + 2\tau_1')\alpha_\omega\tau_{0\omega} \right.   \nonumber \\
&~~~~~~~~~~~~~~~~ \left. - \alpha_\omega[2 \tau_1' \tau_{2\omega} + \eps_F(\tau_1' \tau_{2\omega})'] 
 \right\}\:,
\end{align}
where $\alpha_{\omega} = (2 - \i\omega\tau_1)/(1-\i\omega\tau_1)$,  $\sigma_0 = e^2 n \tau_{1}/m$ is the ``dark'' conductivity of 2DEG, $n = g m \eps_F/(2 \pi \hbar^2)$ is the electron density, and all energy dependent quantities are taken at the Fermi energy $\eps_F$.
Equation~\eqref{gamma_par} can be applied to different scattering mechanisms of 2D electrons characterized by the energy dependence of $\tau_1$ and $\tau_2$. Particularly, it is seen that short-range scattering yielding $\tau_1$ and $\tau_2$ independent of energy does not contribute to the transverse photoconductivity of 2DEG with parabolic spectrum.

\begin{figure}[htpb]
\includegraphics[width=0.49\textwidth]{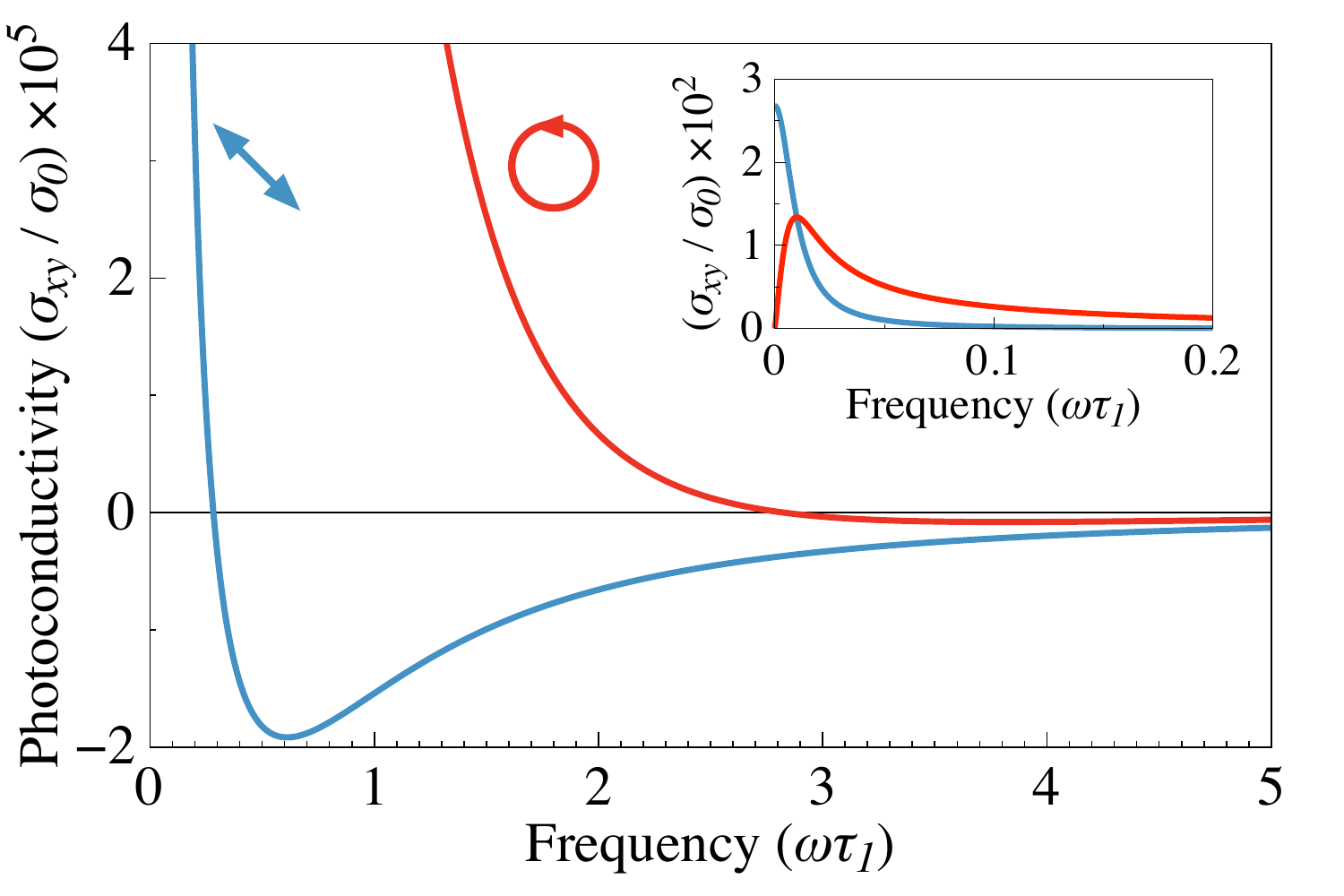}
\caption{\label{fig2} Transverse photoconductivity of 2DEG with parabolic spectrum induced by linearly (blue lines) and circularly (red lines) polarized radiation. The inset shows the behaviour at low frequencies $\omega \tau_0 \lesssim 1$. The curves are calculated after Eq.~\eqref{gamma_par} with  $\tau_1 = 2 \tau_2 \propto \eps$ corresponding to Coulomb scatterers. $I = 1$~W/cm$^2$, $\tau_1 = 1$~ps, $\tau_0 = 100 \tau_1$, $m = 0.03~m_0$, $\eps_F = 50$~meV, $n_\omega = 3$.
}
\end{figure}

Figure~\ref{fig2} shows the photoconductivity $\sigma_{xy} = \gamma_{2,3} |\bm E|^2$ calculated after Eq.~\eqref{gamma_par} with  $\tau_1 = 2 \tau_2 \propto \eps$ corresponding to unscreened Coulomb scatterers. The parameters used for calculations are given in the caption of Fig.~\ref{fig2} and correspond approximately to bi-layer graphene~\cite{Candussio2020}. The curves are calculated for $I = 1$~W/cm$^2$, where $I = c n_\omega |\bm E|^2/2\pi$ is the intensity of radiation, and $n_\omega$ is the refraction index of the medium surrounding 2DEG.
At $\omega \tau_0 \lesssim 1$ the photoconductivity is dominated by the first contribution in Eq.~\eqref{gamma_par} related to the heating mechanism. This contribution is on the order of $\sigma_0 \xi$, where $\xi = e^2 |\bm E|^2 \tau_1 \tau_0/(m\eps_F)$ is the dimensionless parameter. Its value is $\xi \approx 7\times 10^{-3}$ for used parameters.
The heating contribution decays at first within the narrow frequency range $\omega \tau_0 \sim 1$. At $\omega \tau_0 \gg 1$ the heating contribution and the second contribution in Eq.~\eqref{gamma_par} related to optical alignment are comparable. They both decay within a much wider frequency range $\omega \tau_1 \sim 1$, and the photoconductivity is determined by the interplay of both. In this frequency range $\sigma_{xy}/\sigma_0 \sim \xi \tau_1/\tau_0 \approx 7\times 10^{-5}$. Interestingly, interplay of two contributions may result in the change of transverse current sign with increasing frequency, as illustrated in Fig.~\ref{fig2}.

\section{Electron gas with an arbitrary dispersion} \label{arbitrary}

Let us now calculate transverse photoconductivity for an arbitrary energy dispersion of electrons $\eps(|\bm p|)$. We introduce an energy-dependent effective electron mass $m(\eps) = p/v$, where $p = |\bm p|$ and $v = \partial \eps/\partial p$. We start with the general Eq.~\eqref{jy_3} for the current. After calculating the derivatives in the right-hand side of Eq.~\eqref{jy_3} one obtains 
\begin{multline}
\label{jy_4}
j_y = e^2 g F \sum \limits_{\bm p} v_x v_y m \left( \frac{\tau_1}{m} \right)' \bar{f_2} \\
+ e^2 g \sum \limits_{\bm p}  \left[\frac{\tau_1}{m} + \frac{m v^2}{2} \left( \frac{\tau_1}{m} \right)' \right] \left( E_y^* \tilde{f_2}  + E_y \tilde{f_2^*} \right)+  \\
+ \frac{e^2 g}{2} \sum \limits_{\bm p} \left[ 2 v_x v_y E_x^* \tilde{f_2} -  (v_x^2 - v_y^2) E_y^* \tilde{f_2} + \mathrm{c.c.} \right] m \left( \frac{\tau_1}{m} \right)'\:.    
\end{multline} 
To calculate the first contribution in the right-hand side of Eq.~\eqref{jy_4} one should multiply Eq.~\eqref{barf2} for the distribution function  $\bar{f_2}$ by  $v_x v_y m (\tau_1/m)'$ and sum the result over $\bm p$.  The remaining two contributions are calculated in the same way using Eq.~\eqref{tildef2}. Integration  by parts in the derived expressions yields 
%\begin{multline}
%\label{jy_5}
%j_y = -e^3 g F \sum \limits_{\bm p} \tau_2 v_x v_y m \left( \frac{\tau_1}{m} \right)' \left( \bm E^* \cdot \pderiv{\tilde{f_1}}{\bm p} + \bm E \cdot \pderiv{\tilde{f_1}^*}{\bm p} \right) \\
%- e^3 g \tau_{0\omega} \sum \limits_{\bm p} \left[\frac{\tau_1}{m} + \frac{m v^2}{2} \left( \frac{\tau_1}{m} \right)' \right] E_y^* \left(F \pderiv{\tilde{f_1}}{p_x} +  \bm E \cdot \pderiv{\bar{f_1}}{\bm p} \right) \\
%- \frac{e^3 g}{2} \sum \limits_{\bm p} \tau_{2\omega} m \left( \frac{\tau_1}{m} \right)' [2E_x^* v_x v_y -  E_y^* (v_x^2 - v_y^2)]   \left(F \pderiv{\tilde{f_1}}{p_x} +  \bm E \cdot \pderiv{\bar{f_1}}{\bm p} \right)   + \mathrm{c.c.} 
%\end{multline} 
\begin{multline}
\label{jy_5}
j_y = e^3 g F \sum \limits_{\bm p} \tilde{f_1} \bm E^* \cdot \pderiv{}{\bm p} \left[\tau_2 v_x v_y m \left( \frac{\tau_1}{m} \right)'\right]  \\
+ e^3 g \tau_{0\omega} \sum \limits_{\bm p} E_y^* \left( \tilde{f_1} \bm F + \bar{f_1} \bm E \right) \cdot \pderiv{}{\bm p}  \left[\frac{\tau_1}{m} + \frac{m v^2}{2} \left( \frac{\tau_1}{m} \right)' \right]  \\
+  \frac{e^3 g}{2} \sum \limits_{\bm p} \left( \tilde{f_1} \bm F + \bar{f_1} \bm E \right) \cdot \\
\cdot \pderiv{}{\bm p} \left[\tau_{2\omega} m \left( \frac{\tau_1}{m} \right)' [2E_x^* v_x v_y -  E_y^* (v_x^2 - v_y^2)] \right]    + \mathrm{c.c.} 
\end{multline} 
Finally, substitution of Eqs.~\eqref{f1} for $\bar{f_1}$ and $\tilde{f_1}$ in Eq.~\eqref{jy_5}, calculation of derivatives in the right-hand side of Eq.~\eqref{jy_5} and summation over $\bm p$ at $T_0 = 0$ yields
%\begin{multline}
%\label{jy_6}
%j_y = -e^4 g F |\bm E|^2 S_2 \sum \limits_{\bm p} f_0'  v^2 \mathrm{Re} \left\{  \tau_{1\omega} \tau_2 \left( \frac{\tau_1}{m} \right)'  \right. \\
%\left. +  \frac{m^2v^2\tau_{1\omega}}{4} \left[ \frac{\tau_2}{m} \left( \frac{\tau_1}{m} \right)' \right]'  \right. \\
%\left. + \frac{\tau_{0\omega} (\tau_{1\omega} + \tau_1)}{2}\left[\frac{\tau_1}{m} + \frac{m v^2}{2} \left( \frac{\tau_1}{m} \right)' \right]'   \right\} \\
%+ e^4 g F |\bm E|^2 S_3  \sum \limits_{\bm p} f_0' v^2 \mathrm{Im} \left\{ (\tau_1 + \tau_{1\omega}) \tau_{2\omega} \left( \frac{\tau_1}{m} \right)'  \right. \\
%\left.  +  \frac{m^2v^2 (\tau_1 + \tau_{1\omega})}{4} \left[ \frac{\tau_{2\omega}}{m} \left( \frac{\tau_1}{m} \right)' \right]' \right. \\
%\left. - \frac{\tau_{0\omega} (\tau_{1\omega} + \tau_1)}{2}\left[\frac{\tau_1}{m} + \frac{m v^2}{2} \left( \frac{\tau_1}{m} \right)' \right]'
%\right \}  \:. 
%\end{multline} 
\begin{align}
\label{gamma_arbitrary}
\gamma_2 &= \sigma_0 e^2  \mathrm{Re} \left\{ \alpha_\omega \tau_{0\omega}\left[\frac{\tau_1}{m} + \frac{m v^2}{2} \left( \frac{\tau_1}{m} \right)' \right]'   \right. \nonumber \\
&\left. +  \frac{m^2v^2}{2(1-\i\omega\tau_1)} \left[ \frac{\tau_2}{m} \left( \frac{\tau_1}{m} \right)' \right]'  +  \frac{2\tau_2}{1-\i\omega\tau_1} \left( \frac{\tau_1}{m} \right)'  \right\} \:, \nonumber \\
\gamma_3 &= \sigma_0 e^2   \mathrm{Im} \left\{ \alpha_\omega \tau_{0\omega} \left[\frac{\tau_1}{m} + \frac{m v^2}{2} \left( \frac{\tau_1}{m} \right)' \right]'
 \right. \nonumber \\
&\left. -  \frac{\alpha_\omega m^2v^2}{2} \left[ \frac{\tau_{2\omega}}{m} \left( \frac{\tau_1}{m} \right)' \right]' - 2\alpha_\omega\tau_{2\omega} \left( \frac{\tau_1}{m} \right)'
\right \}  \:,
\end{align} 
where $\sigma_0 = e^2 n \tau_{1}/m$ is the ``dark'' conductivity of 2DEG, and energy dependent quantities are taken at $\eps_F$.

Equation~\eqref{gamma_arbitrary} is general and can be used to calculate anisotropic photoconductivity of 2DEG with arbitrary energy dispersion $\eps(p)$ and for arbitrary energy dependence of relaxation times. For parabolic dispersion, $m$ is energy-independent, $mv^2/2 = \eps$, and Eq.~\eqref{gamma_arbitrary} is reduced to Eq.~\eqref{gamma_par} obtained in Sec.~\ref{parabolic}. For linear dispersion $\eps = v_0  p$, relevant to graphene, one has $m = \eps/v_0^2$ and $mv^2 = \eps$. For $\tau_1 = 2\tau_2 \propto \eps^{-1}$ corresponding to scattering at the short-range potential in graphene  Eq.~\eqref{gamma_arbitrary} yields
\begin{equation}
\label{gamma_graphene}
\gamma_2^{(\delta)} = \frac{\sigma_0 e^2 v_0^2 \tau_1^2}{2\eps_F^2(1+\omega^2\tau_1^2)}\:,~~~ \gamma_3^{(\delta)} = - \frac{12 \omega \tau_1}{4+\omega^2\tau_1^2} \gamma_2  \:.
\end{equation}

\begin{figure}[htpb]
\includegraphics[width=0.49\textwidth]{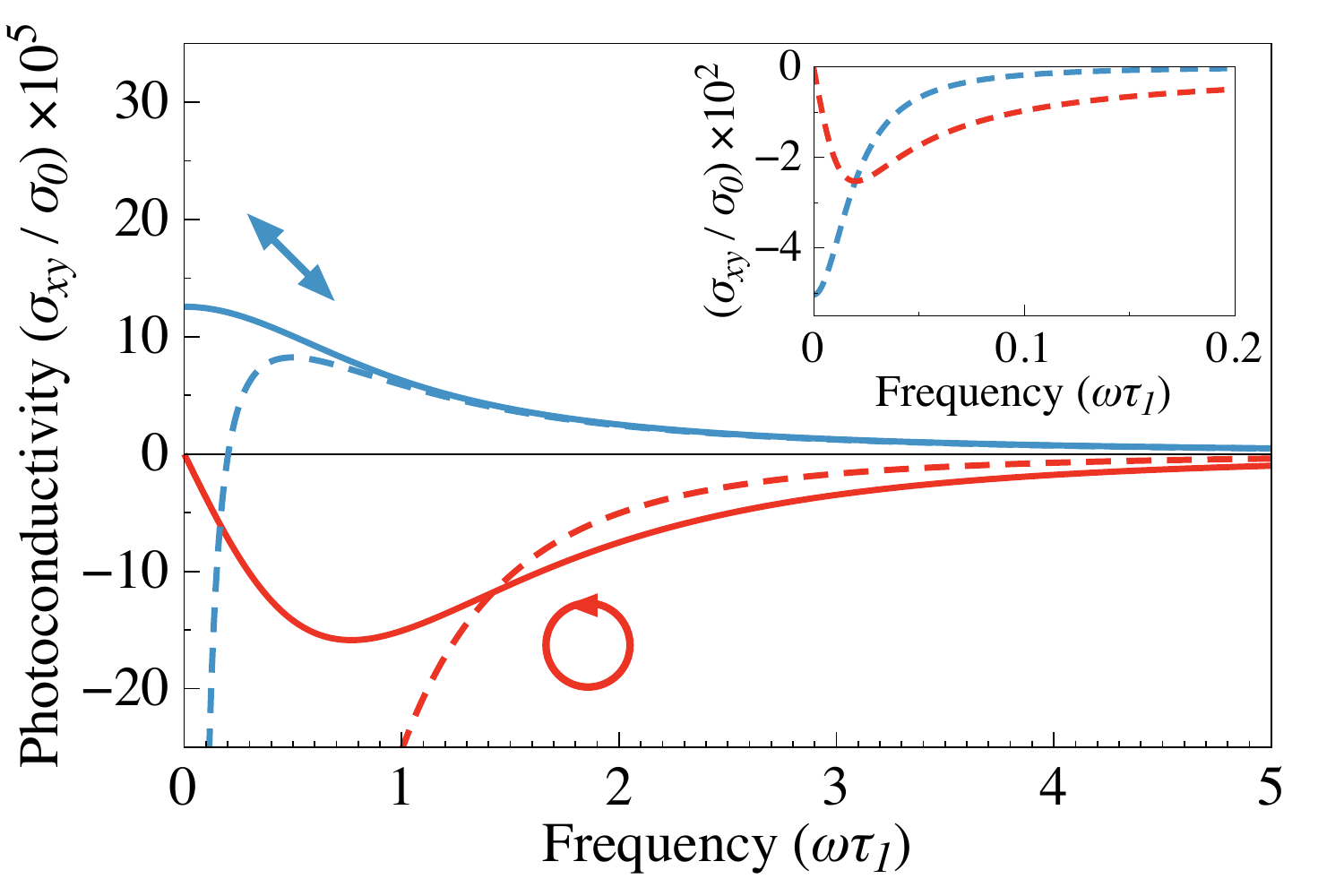}
\caption{\label{fig3}
Transverse photoconductivity of 2DEG in graphene induced by linearly (blue lines) and circularly (red lines) polarized radiation. The inset shows the behaviour at low frequencies $\omega \tau_0 \lesssim 1$. The solid curves are calculated after Eq.~\eqref{gamma_graphene} with  $\tau_1 = 2 \tau_2 \propto \eps^{-1}$ corresponding to short-range scatterers. The dashed curves are calculated after Eq.~\eqref{gamma_arbitrary} with  $\tau_1 = 2 \tau_2 \propto \eps/(\eps^2 + \eps_0^2)$ corresponding to mixture of Coulomb and short-range scatterers. $I = 1$~W/cm$^2$, $\tau_1 = 1$~ps, $\tau_0 = 100 \tau_1$, $\eps_F = \eps_0 = 50$~meV, $v_0 = 10^8$~cm/s, $n_\omega = 3$.
}
\end{figure}

Figure~\ref{fig3} shows the photoconductivity $\sigma_{xy} = \gamma_{2,3} |\bm E|^2$ in graphene calculated after Eq.~\eqref{gamma_graphene} for short-range scatterers and after Eq.~\eqref{gamma_arbitrary} with  $\tau_1 = 2 \tau_2 \propto \eps/(\eps^2 + \eps_0^2)$ corresponding to mixture of Coulomb and short-range scatterers, $\eps_0$ is a parameter.
Note that the heating contribution vanishes in graphene with short-range scatterers, and hence the  photoconductivity~\eqref{gamma_graphene} is independent of $\tau_0$. However, the addition of Coulomb centers restores the heating contribution and significantly enhances the photoconductivity at $\omega \tau_0 \lesssim 1$. Note that scattering solely at Coulomb centers yields $\tau_{1,2} \propto \eps$ resulting in $\sigma_{xy} = 0$. The values of $\sigma_{xy}/\sigma_0$ at $\omega \tau_0 \gg 1$ are about an order of magnitude larger in monolayer graphene than in bi-layer graphene, see Fig.~\ref{fig2}, due to the smaller effective mass of electrons in graphene.

%\begin{multline}
%\label{jy_7}
%\sigma_{xy} = e^2 |\bm E|^2 S_2 \mathrm{Re} \sigma \left\{ 2\tau_2  \left( \frac{\tau_1}{m} \right)' + \frac{m^2v_F^2}{2} \left[ \frac{\tau_2}{m} \left( \frac{\tau_1}{m} \right)' \right]'  \right. \\
%\left. + \frac{\tau_0(2-\omega^2\tau_0\tau_1)}{1+\omega^2\tau_0^2} \left[\frac{\tau_1}{m} + \frac{m v^2}{2} \left( \frac{\tau_1}{m} \right)' \right]' \right\}
%\\
%- \frac{2 e^2  |\bm E|^2 S_3 \omega \mathrm{Re} \sigma }{1 + \omega^2 \tau_2^2} \\
%\times \left\{ \tau_2(\tau_1+2\tau_2+\omega^2\tau_1^2\tau_2) \left[ \left( \frac{\tau_1}{m} \right)' + \frac{m^2v_F^2}{4} \left[ \frac{1}{m} \left( \frac{\tau_1}{m} \right)' \right]'  \right] \right. \\
%\left. + \frac{mv_F^2[\tau_1+4\tau_2+\omega^2\tau_1\tau_2(2\tau_1-\tau_2)]}{4(1 + \omega^2 \tau_2^2)} \left( \frac{\tau_1}{m} \right)' \tau_2' \right. \\
%\left. - \frac{\omega \tau_0^2 (2+\omega^2 \tau_1^2)}{2(1+\omega^2\tau_0^2)} \left[\frac{\tau_1}{m} + \frac{m v^2}{2} \left( \frac{\tau_1}{m} \right)' \right]'
%\right \}  \:. 
%\end{multline} 

\section{Role of finite temperature and thermalization} \label{thermal}

In this section we calculate the transverse photoconductivity of 2DEG at finite temperature. At finite temperature the distribution function of 2DEG is smeared over energy and thermalization process caused by electron-electron collisions affects the kinetics of the zeroth angular harmonic $\aver{f}$ of the distribution function. We describe this kinetics by the collision integral~\eqref{St_zero}.   
We assume that, in the absence of external fields $\bm F = 0$, $\bm E = 0$, the temperature of 2DEG is $T_0$ and its chemical potential $\mu_0$. Heating or cooling of the gas by combined action of incident radiation and static electric field and fast electron-electron collisions result in a different temperature $T$ and chemical potential $\mu$ of electron subsystem. Further, we discuss the thermalization of the oscillating correction $\tilde{f_2} \exp(-\i\omega t) + \mathrm{c.c.}$ to the distribution function of 2D electrons, since this correction determines the heating contribution to the current in Eq.~\eqref{jy_4}. Therefore, $T$ and $\mu$ also oscillate with frequency $\omega$.

 To determine $T$ one should calculate the change of  the total electron energy $\Delta E(t)$ at a given moment of time and consider how it is redistributed between thermalized electrons. The change of the total electron energy is 
\begin{equation}
\Delta E = \sum \limits_{\bm p} \eps (f - f_0) = \sum \limits_{\bm p} \eps [f_{T}(\mu,T) - f_0]\:,
\end{equation}
because $\sum_{\bm p} \eps f = \sum_{\bm p} \eps f_{T}(\mu,T)$. Assuming $\Delta T/T_0 \ll 1$, where $\Delta T = T - T_0$, one has
\begin{equation}
\label{dE}
\Delta E \approx \frac{\Delta T \sum_{\bm p} \eps A}{T_0}\:,
\end{equation}
where 
\begin{equation}
 A(\eps) = - \left[\eps - \mu_0 + T_0 \frac{d \mu}{d T}(T_0)  \right] f_0'\:,
\end{equation}
and $ d \mu/d T$ can be found from the particle conservation constraint $\sum_{\bm p} [f_T(\mu,T) - f_0] = 0$. On the other hand, $\Delta E$ is found from Eq.~\eqref{tildef2} bearing in mind that thermalization conserves the total energy of 2DEG. Taking into account Eq.~\eqref{dE} one thus finds
\begin{equation}
\label{dT}
\frac{\Delta T}{T_0} =  \frac{\tau_{0\omega} \sum_{\bm p} G \eps}{\sum_{\bm p} \eps A} \e^{-\i\omega t} + \mathrm{c.c.}\:,
\end{equation} 
where
\begin{equation}
\label{generation}
G = -e \aver{\bm F \cdot \pderiv{\tilde{f_1}}{\bm p} +  \bm E \cdot \pderiv{\bar{f_1}}{\bm p}}\:.
\end{equation}

Solution of the kinetic equation~\eqref{tildef2} with the collision integral~\eqref{St_zero} and $\Delta T$ given by Eq.~\eqref{dT} yields
\begin{equation}
\label{f2aver}
\aver{\tilde{f_2}} = \frac{\tau_{0\omega} }{\tau_{0\omega} + \tau_{ee}} \left( G \tau_{ee} +A \frac{  \tau_{0\omega}\sum_{\bm p} G \eps}{\sum_{\bm p} A \eps} \right)\:.
\end{equation}
Equation~\eqref{f2aver} is then used to calculate the heating contribution to the electric current in Eq.~\eqref{jy_4}.  
%Using Eqs.~\eqref{f2aver} and \eqref{f1} one obtains
%\begin{equation}
%\label{sumG2}
%\sum \limits_{\bm p} G \tilde{f_2} = -\frac{e^2 \tau_{0\omega} F E_x}{2}
% \sum \limits_{\bm p} v^2 (\tau_1 + \tau_{1\omega}) f_0' H \:,
%\end{equation}
%where
Computation analogous to the one in Sec.~\ref{arbitrary} yields
\begin{align}
\label{gamma_thermal}
\gamma_2 &= \sigma_0 e^2 \mathrm{Re} \left\langle \alpha_\omega \tau_{0\omega} H  + \frac{2\tau_2}{1-\i\omega\tau_1} \left( \frac{\tau_1}{m} \right)'    \right. \nonumber \\
& \left.  +  \frac{m^2v^2}{2(1-\i\omega\tau_1)} \left[ \frac{\tau_2}{m} \left( \frac{\tau_1}{m} \right)' \right]'  \right\rangle_\eps \:,  \nonumber \\
 \gamma_3 &= \sigma_0 e^2  \mathrm{Im} \left\langle \alpha_\omega \tau_{0\omega} H - 2\alpha_\omega\tau_{2\omega} \left( \frac{\tau_1}{m} \right)'
 \right. \nonumber \\
& \left. -  \frac{\alpha_\omega m^2v^2}{2} \left[ \frac{\tau_{2\omega}}{m} \left( \frac{\tau_1}{m} \right)' \right]' 
\right \rangle_{\eps}  \:.
\end{align} 
Here $H$ is given by 
\begin{equation}
\label{H}
H =  \frac{\tau_{ee}}{\tau_{0\omega} + \tau_{ee}} Q'  + \frac{\tau_{0\omega}}{\tau_{0\omega} + \tau_{ee}} \frac{\sum_{\bm p} A Q}{\sum_{\bm p} A \eps} \:
\end{equation}
with
\begin{equation}
Q = \frac{\tau_1}{m} + \frac{m v^2}{2} \left( \frac{\tau_1}{m} \right)' \:,
\end{equation}
and the energy averaging $\aver{\dots}_\eps$ is defined as
\begin{equation}
\aver{...}_\eps = \frac{\sum_{\bm p} (...) v^2 \tau_1 f_0'}{\sum_{\bm p} v^2 \tau_1 f_0'}\:.
\end{equation}

\begin{figure}[htpb]
\includegraphics[width=0.49\textwidth]{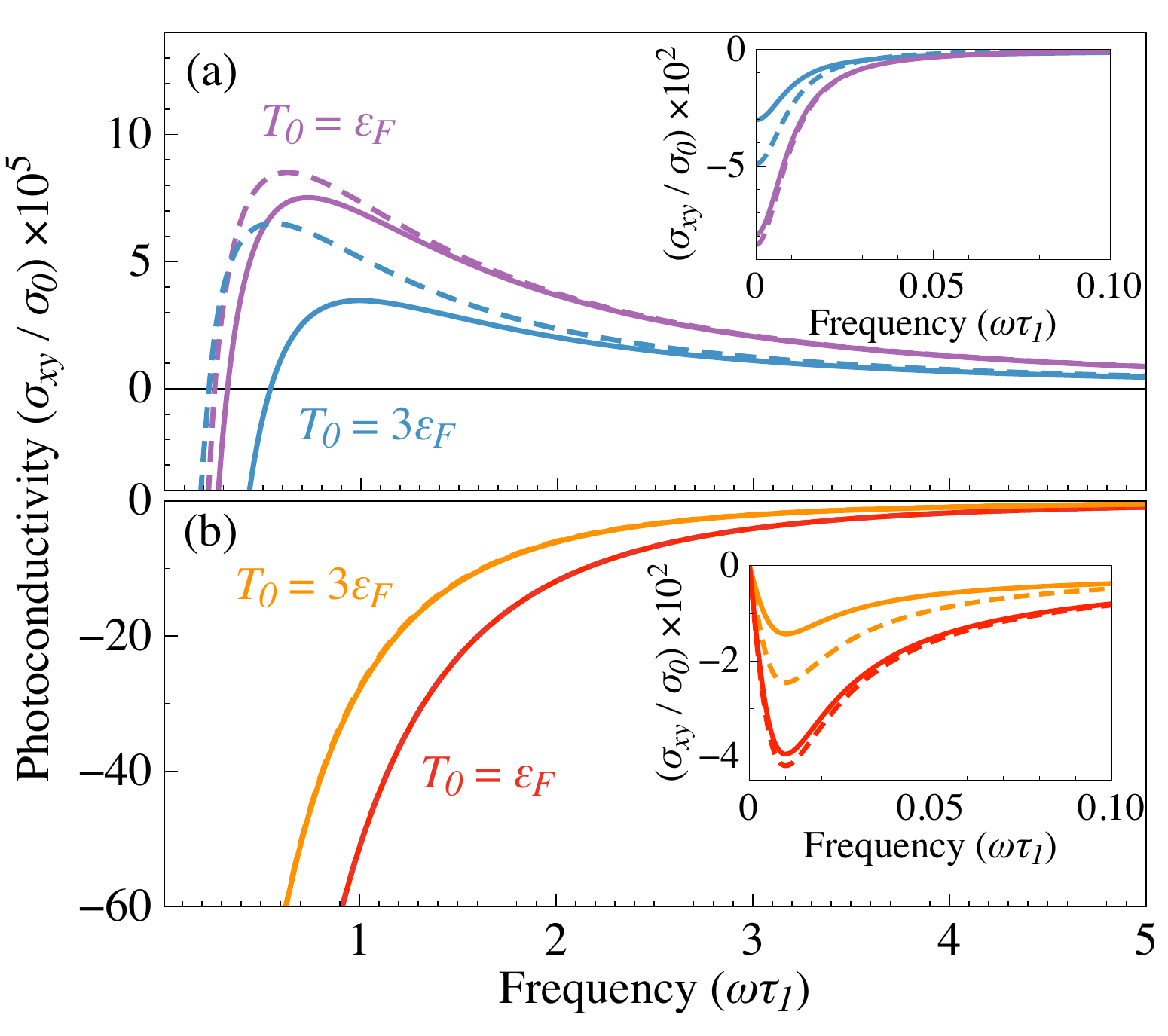}
\caption{\label{fig4}
Transverse photoconductivity of 2DEG in graphene at nonzero temperature for linearly (a) and circularly (b) polarized radiation. The inset shows the behaviour at low frequencies $\omega \tau_0 \lesssim 1$. The  curves are calculated after Eq.~\eqref{gamma_thermal} with $\tau_{ee} = 10\tau_1$ (solid) and $\tau_{ee} = +\infty$ (dashed). $\tau_1 = 2 \tau_2 \propto \eps/(\eps^2 + \eps_0^2)$, $\eps_0 = 50$~meV, $\eps_F = 10$~meV,  $I = 1$~W/cm$^2$, $\tau_1(\eps_0) = 1$~ps, $\tau_0 = 100 \tau_1$, $v_0 = 10^8$~cm/s, $n_\omega = 3$.
}
\end{figure}

Equation~\eqref{gamma_thermal} allows one to calculate the transverse photoconductivity of 2DEG at a given temperature $T_0$. At $T_0 = 0$, one has $\sum_{\bm p} A Q/\sum_{\bm p} A \eps = Q' (\mu_0)$ and, as follows from Eq.~\eqref{H}, $H (\mu_0) = Q'(\mu_0)$.  In that case, thermalization does not affect the photoconductivity, and $\gamma_{2,3}$ coincide with those given by Eq.~\eqref{gamma_arbitrary}.
 Figure~\ref{fig4} presents the transverse photoconductivity $\sigma_{xy} = \gamma_{2,3} |\bm E|^2$ calculated after Eq.~\eqref{gamma_thermal} for graphene. Calculations are done for a fixed electron density defined by the Fermi energy $\eps_F = 10$~meV, and two temperatures $T_0 = \eps_F$ and $T_0 = 3 \eps_F$. The corresponding values of the chemical potential are $\mu_0 \approx -5.7$~meV and $\mu_0 \approx -86$~meV, respectively. The latter case corresponds to the Boltzmann distribution of electrons.
Thermalization of electrons starts to play a role at $T_0 \sim \eps_F$. As blue and red curves in Fig.~\ref{fig4} show, both circular and linear photoconductivities are suppressed by thermalization at $\omega \tau_{ee} \lesssim 1$, and this suppression becomes stronger with increase of temperature. However, such a suppression is not a general behaviour. Thermalization aims to redistribute electrons over energy and it may cause either suppression or enhancement of the photoconductivity depending on how electron velocity and relaxation time $\tau_1$ depend on energy. For instance, in the case of parabolic dispersion and Coulomb scatterers shown in Fig.~\ref{fig2}, the thermalization does not affect $\sigma_{xy}$ at all, because in that case $H$ given by Eq.~\eqref{H} is independent of energy.

\section{Summary} \label{conclusion}

To summarize, we have studied the transverse photoconductivity of 2DEG caused by intraband absorption of circularly and linearly polarized radiation. The transverse dc current has two contributions: (i) due to the optical alignment of electron momenta and (ii) due to the dynamic heating and cooling of 2DEG. The heating contribution is dominant at low frequencies $\omega \tau_0 \lesssim 1$, where $\tau_0$ is the energy relaxation time. In this range, the transverse photoconductivity reaches $\sim 1$~\% of  the ``dark'' conductivity of 2DEG at 1 W cm$^{-2}$ of the radiation intensity. At higher frequencies, the transverse current is determined by the relaxation of the first and second angular harmonics of the distribution function. We have developed the microscopic theory of transverse photoconductivity for arbitrary electron spectrum and scattering mechanism. The value and the sign of the calculated photoconductivity of 2DEG with parabolic and linear energy dispersion significantly depend on the scattering mechanism. Further, we have shown that thermalization processes caused by electron-electron collisions have a negligible impact on transverse photoconductivity of a nearly degenerate 2DEG, but may contribute considerably at higher temperatures, when the Boltzmann distribution is formed.

\acknowledgments
The author thanks S. A. Tarasenko and M. M. Glazov for fruitful discussions. The work was financially supported by the Basis Foundation for the Advancement of Theoretical Physics and Mathematics.

\end{document}